\documentclass[prd,preprint,showpacs,preprintnumbers,nofootinbib,eqsecnum,superscriptaddress]{revtex4}

 \usepackage[dvips,final]{graphicx}
  \usepackage{amssymb}
   \usepackage{amsmath}
    \usepackage{amsfonts}
     \usepackage{epsfig}
      \usepackage{bm}

\usepackage{mathpazo}

\usepackage[section]{placeins}

\usepackage{multirow}
\usepackage{ctable}
\usepackage{booktabs}
\usepackage{array}
\usepackage{tabularx}
\usepackage{xcolor}
\usepackage{pstricks}

\begin{document}

\title{Open charm meson production at BNL RHIC within $\bm{k_{t}}$-factorization approach and revision of their semileptonic decays \\}

\author{Rafa{\l} Maciu{\l}a}
\email{rafal.maciula@ifj.edu.pl} 
\affiliation{Institute of Nuclear Physics PAN, PL-31-342 Cracow, Poland}

\author{Antoni Szczurek\footnote{also at University of Rzesz\'ow, PL-35-959 Rzesz\'ow, Poland}}
\email{antoni.szczurek@ifj.edu.pl}
\affiliation{Institute of Nuclear Physics PAN, PL-31-342 Cracow, Poland}

\author{Marta {\L}uszczak}
\email{luszczak@univ.rzeszow.pl} 
\affiliation{University of Rzesz\'ow, PL-35-959 Rzesz\'ow, Poland}

\date{\today}

\begin{abstract}
We discuss inclusive production of open charm mesons in proton-proton scattering at the BNL RHIC. The calculation is performed in the framework of $k_t$-factorization approach which effectively includes higher-order pQCD corrections. Different models of unintegrated gluon distributions (UGDF) from the literature are used. We focus on UGDF models favoured by the LHC data and on a new up-to-date parametrizations based on the HERA collider DIS high-precision data. Results of the $k_t$-factorization approach are compared to next-to-leading order collinear predictions. The hadronization of heavy quarks is done by means of fragmentation function technique. The theoretical transverse momentum distributions of charmed mesons are compared with recent experimental data of the STAR collaboration at $\sqrt{s} = 200$ and $500$ GeV. Theoretical uncertainties related to the choice of renormalization and factorization scales as well as due to the quark mass are discussed. Very good description of the measured integrated cross sections and differential distributions is obtained for the Jung setB$0$ CCFM UGDF. Revised charm and bottom theoretical cross sections corresponding to those measured recently by the STAR and PHENIX collaborations for semileptonic decays of $D$ and $B$ mesons are presented. Significant improvement in theoretical description of the non-photonic electrons measurements is clearly obtained with respect to the previous studies within the $k_{t}$-factorization.

\end{abstract}

\pacs{13.87.Ce, 14.65.Dw}

\maketitle

\section{Introduction}
Production of heavy quarks is of substantial and ongoing interest in high energy hadronic collisions. This statement had not changed in nearly 40 years, when the charm and bottom flavoured particles were discovered. The energy scale for the production of charm and bottom quarks is significantly higher than the typical Quantum Chromodynamics (QCD) scale, $\Lambda_{QCD}\sim 0.2$ GeV. This gives the value of strong coupling of the order of $\alpha_S \sim 0.2 - 0.3$, which is small enough to apply perturbative QCD techniques. At the present level of knowledge and current experimental abilities, heavy flavours are known as one of the best testing grounds of the theory of hard QCD interactions. Theoretical analyses of charm and bottom cross sections in proton-proton interactions provide
an unique precision tool in this context. Due to their mass, heavy quarks are also belived to be a special probe of the medium created in heavy ion collisions.
Since heavy quarks are only produced in the initial stage of the heavy ion collisions, heavy quark distributions from proton-proton interactions
supply a well defined initial state. Their further propagation through the hot and dense medium probes its interesting characteristics.
 
Measurements of charm and bottom cross sections at hadron colliders can be performed in the so-called indirect way.
This method is based on measurement of leptons from semileptonic decays of open charm and bottom mesons, which are often called non-photonic. The semileptonic decay modes allow for an indirect measurement of the D and B meson cross sections and by further extrapolation the charm and bottom quark cross sections. The indirect methodologies to measure non-photonic electrons/muons have been used since the early 1970's \cite{Appel:1974yu}.

The decay of hadrons by the weak interaction can be viewed as a process of decay of their constituent quarks. 
The charm and bottom flavours are not preserved in weak interactions, so their weak decays are possible. Within the semileptonic decays, the allowed (according to electric charge conservation) transitions $b \to c,u$ and $c \to s,d$ are involved by the emission of charged $W$ boson, which further creates a lepton and the corresponding antineutrino $W \to l \bar{\nu}_l$. The semileptonic decay widths are proportional to the square of the appropriate elements of the Cabbibo-Kobayashi-Maskawa (CKM) quark-mixing matrix \cite{Kobayashi:1973fv}, which contains information on the strength of flavour-changing
weak decays. The decays within the same quark generation are strongly favoured over decays between generations.

Main virtues of the semileptonic modes come from the fact that they have bigger branching fractions than the hadronic $D$ meson decay channels. Moreover, in the former the effects of strong interactions can be isolated and thus they are better accessible
experimentally. In addition, direct lepton production through weak and electromagnetic interactions in hadronic collisions is suppressed relative to the strong interaction by twelve and four orders of magnitude, respectively \cite{Halzen:1984mc}.

However, electrons and muons are certainly not rare paticles because they are abundantly produced in light hadron decays.
These rather problematic background must be accounted for and eliminated through experimental techniques. If the primary sources of background are well understood and substracted, the remaining events can then be attributed to the heavy flavour signal.

Another method for experimental investigation of charm and bottom quark production at hadron colliders is the direct procedure
based on full reconstruction of all decay products of open charm and bottom mesons, for instance in the 
$D^0 \to K^- \pi^+ $, $D^+ \to K^- \pi^+ \pi^+$ or $B^+ \to J/\psi K^+ \to K^+ \mu^+ \mu^-$ channels. The hadronic decay products can be used to built
invariant mass distributions, permitting direct observation of $D$ or $B$ meson as a peak in the experimental invariant mass spectrum. Open charm $D$ and $B$ mesons are characterized by rather long lifetimes, of the order of $\sim 10^{-13}$ and $\sim 10^{-12}$ seconds, respecitvely. Charm and bottom quarks decay essentially at the collision vertex, while heavy flavour mesons decay from a secondary vertex offset by the boosted lifetime of the paticle. In the direct approach the charm and bottom contributions can be well separated, which is not the case in the indirect method. In the latter case, it can be achieved only within the analysis of lepton-meson (e.g. $e$-$D$) correlations \cite{Mischke:2008af}, which are easily available e.g. in the $k_t$-factorization approach.
 
The STAR and PHENIX collaborations have measured transverse momentum distributions of electrons coming from the semileptonic decays of
heavy flavoured hadrons in proton-proton scattering at the RHIC energy $\sqrt{s} = 200$ GeV with lepton transverse momenta up to $10$ GeV in the midrapidity region \cite{Adare:2006hc,Abelev:2006db}. In addition, the STAR collaboration was able to separate the charm and bottom contributions to the spectra of heavy flavour electrons \cite{Agakishiev:2011mr}.
The PHENIX collaboration has also measured non-photonic dilepton invariant mass spectrum from $0$ to $8$ GeV in proton-proton collisions at $\sqrt{s}=200$ GeV \cite{Adare:2008ac}. On the theoretical side, the cross sections for inclusive production of the non-photonic electrons at RHIC have been studied theoretically up to the next-to-leading order pQCD collinear approximation within the FONLL approach in Ref.~\cite{Cacciari:2005rk}. The first theoretical investigation within the competitive QCD $k_t$-factorization framework was done in Refs.~\cite{Luszczak:2008je,Maciula:2010yw}, including more exclusive studies of kinematical correlations.

Very recently, the STAR collaboration measured for the first time transverse momentum spectra of $D^*$ and $D^0$ mesons up to $6$ GeV at $\sqrt{s} = 200$ \cite{Adamczyk:2012af} and $500$ GeV \cite{STAR-Dmeson500} in proton-proton collisions. Before, studies of charm production at RHIC through hadronic decay channels were performed only in Cu-Cu collisions \cite{Baumgart:2008zz} and in proton-proton scattering but for $D^*$ mesons produced in jets \cite{Abelev:2009aj}.
Up to now, the new STAR proton-proton data on open charm production were studied only in the context of high energy pA collisions in the Color Glass Condensate framework with the unintegrated gluon densities from the solution of rcBK equation \cite{Fujii:2013yja}.   

Our aim here is to make first theoretical analysis of the measured hadron-level charm differential cross section within the $k_t$-factorization approach. The open charm meson data allow us to make more direct comparison of the theoretical predictions and RHIC experimental results on heavy flavour production without including additional step related to the semileptonic decays. Recently, the formalism of the $k_t$-factorization approach has been found to give very good description of open charm \cite{Maciula:2013wg} and bottom \cite{Jung:2011yt,Karpishkov:2014epa} production rates and kinematical correlations in proton-proton scattering at $\sqrt{s} = 7$ TeV measured by the ALICE, ATLAS, CMS and LHCb experiments. However, a significant sensitivity of theoretical predictions on the model of unintegrated (transverse momentum dependent) gluon distributions (UGDFs) used in calculations has been also reported.  

Therefore, it is very interesting to make similar study for the STAR experimental data on open heavy flavour production at $\sqrt{s} = 200$ and $500$ GeV. This         
may be a good test of different models of UGDFs in the RHIC kinematical regime where one can probe parton (gluon) distributons at intermediate longitudinal momentum fractions $x_1/x_2 \sim 10^{-2} - 10^{-1}$. Here, we wish to pay particular attention on UGDF models favoured by the LHC data and on a new up-to-date parametrizations based on the HERA collider DIS high-precision data. The present study is an important extension of our previous paper, where charm and bottom cross section at RHIC has been considered in the context of semileptonic decays of open heavy mesons \cite{Luszczak:2008je}.
 
Precise predictions for charmed mesons may also shed new light on the theoretical understanding of non-photonic lepton production at RHIC.
Our previous studies of these processes within the $k_t$-factorization approach were based on rather older models of UGDFs which may be the reason of the reported missing strength in description of the RHIC experimental data. Similar problem was also noticed within the NLO collinear calculations in the FONLL model, where only upper limits of the theoretical predictions are consistent with the relevant STAR and PHENIX data \cite{Cacciari:2005rk}. Therefore, in the following paper we will also revise theoretical cross sections of the non-photonic lepton production at RHIC wihtin the $k_t$-factorizaton approach, taking as a reference point the results obtained in the analysis of the new hadron-level STAR data. In the following calculation, except of updated UGDFs, we also take into account the effect of transformation of semileptonic decay functions between laboratory ($e^{+}e^{-}$ center-of-mass system) and rest frames of decaying $D$ or $B$ mesons.

\section{Theoretical formalism}

Several different mechanisms play a role in heavy quark hadroproduction. In general, in the framework of QCD there are two types of the $\mathcal{O}(\alpha_S^2)$ leading-order (LO) $2 \to 2$ subprocesses: $q\bar q \to Q \overline Q$ and $gg \to Q \overline Q$ \cite{Combridge:1978kx}, often referred to as heavy quark-antiquark pair creation. The first mechanism, $q\bar q$-annihilation, is important only near the threshold and at very large invariant masses of $Q\overline Q$ system or extremely forward rapidities. This contribution is therefore especially important in the case of top quark production, however, for charm and bottom production, starting from RHIC, through Tevatron, up to the LHC it can be safely neglected. At high energies, production of charm and bottom flavours is dominated by the gluon-gluon fusion, which is the starting point of the following analysis. In the case of heavy quark production the $\mathcal{O}(\alpha_S^3)$ next-to-leading order (NLO) perturbative contributions have been found to be of special importance (see e.g. \cite{Beenakker:1990maa}) and have to be included in order to describe the heavy flavour high energy experimental data.

The QED contributions, with one or two photons initiated reactions, such as $\gamma g \to Q \overline Q$, $g \gamma \to Q \overline Q$, $\gamma \gamma \to Q \overline Q$ have been carefully studied in Ref.~\cite{Luszczak:2011uh} together with other sub-leading contributions to production of charm and were found to be negligibly small at high energies.

In the following, the cross sections for charm and bottom quark production in proton-proton collisions are calculated in the framework of the $k_t$-factorization approach. This framework has been successfully applied for different high energy processes, including heavy quark production (see e.g.~\cite{Maciula:2013wg} and references therein). According to this approach, the transverse momenta $k_{t}$'s (virtualities) of partons which initiate reaction are taken into account and the sum of transverse momenta of the final $Q$ and $\overline Q$ no longer
cancels. Then the LO differential cross section for the $Q \overline Q$ pair production reads:
\begin{eqnarray}\label{LO_kt-factorization} 
\frac{d \sigma(h_1 h_2 \to Q \overline Q \, X)}{d y_1 d y_2 d^2p_{1,t} d^2p_{2,t}} &=& \sum_{i,j} \;
\int \frac{d^2 k_{1,t}}{\pi} \frac{d^2 k_{2,t}}{\pi}
\frac{1}{16 \pi^2 (x_1 x_2 s)^2} \; \overline{ | {\cal M}^{off-shell}_{g^* g^* \to Q \overline Q} |^2}
 \\  
&& \times  \; \delta^{2} \left( \vec{k}_{1,t} + \vec{k}_{2,t} 
                 - \vec{p}_{1,t} - \vec{p}_{2,t} \right) \;
{\cal F}_g(x_1,k_{1,t}^2) \; {\cal F}_g(x_2,k_{2,t}^2) \; \nonumber ,   
\end{eqnarray}
where ${\cal F}_g(x_1,k_{1,t}^2)$ and ${\cal F}_g(x_2,k_{2,t}^2)$
are the UGDFs for the both colliding hadrons. The extra integration is over transverse momenta of the initial
partons. Explicit treatment of the transverse part of momenta makes the approach very efficient in studies of correlation observables. The two-dimensional Dirac delta function assures momentum conservation.
The unintegrated (transverse momentum dependent) gluon distributions must be evaluated at:
\begin{equation}
x_1 = \frac{m_{1,t}}{\sqrt{s}}\exp( y_1) 
     + \frac{m_{2,t}}{\sqrt{s}}\exp( y_2), \;\;\;\;\;\;
x_2 = \frac{m_{1,t}}{\sqrt{s}}\exp(-y_1) 
     + \frac{m_{2,t}}{\sqrt{s}}\exp(-y_2), \nonumber
\end{equation}
where $m_{i,t} = \sqrt{p_{i,t}^2 + m_Q^2}$ is the quark/antiquark transverse mass. In the case of heavy quark production at RHIC energies, especially in the central rapidity region, one test kinematical regime of $x > 10^{-2}$.  

The LO matrix element squared for off-shell gluons is taken in the analytic form proposed by Catani, Ciafaloni and Hautmann (CCH) \cite{Catani:1990eg}. This analytic formula was basically derived within the standard QCD framework and can be adopted to the numerical analyses. It was also checked that the CCH expression is consistent with those presented later in Refs.~\cite{Collins:1991ty,Ball:2001pq} and in the limit of $k_{1,t}^2 \to 0$, $k_{2,t}^2 \to 0$ it converges to the on-shell formula.

The calculation of higher-order corrections in the $k_t$-factorization is much more complicated than in the case of collinear approximation.
However, the common statement is that actually in the $k_{t}$-factorization approach with LO off-shell matrix elements some part of higher-order corrections is effectively included. This is due to emission of extra gluons encoded
in the unintegrated gluon densities. More details of the theoretical formalism adopted here can be found in Ref.\cite{Maciula:2013wg}. 
  
In the numerical calculation below we have applied several unintegrated gluon densities which are based on different theoretical assumptions. The Kimber-Martin-Ryskin (KMR) UGDF is derived from a modified DGLAP-BFKL evolution equation \cite{Kimber:2001sc,Watt:2003mx} and has been found recently to work very well in the case of charm and bottom production at the LHC. A special emphasis here is put on the UGDF models obtained as a solution of CCFM evolution equation. Here we use an older Jung set$B0$ parametrization \cite{Jung:2004gs} and up-to-date JH2013 distributions \cite{Hautmann:2013tba} determined from the fits to HERA high-precision DIS measurements. The JH2013 set1 is obtained from the fit to inclusive $F_{2}$ data only while JH2013 set2 is derived from the fit to both $F^{charm}_{2}$ and $F_{2}$ data. The UGDFs based on BFKL and BK equations are not applied in the following analysis since they are dedicated to smaller-$x$ values.    

In the calculation of charm and bottom quark cross sections the central value of numerical results is obtained with the renormalization and factorization scales $\mu^2 = \mu_{R}^{2} = \mu_{F}^{2} = \frac{m^{2}_{1,t} + m^{2}_{2,t}}{2}$ and quark mass $m_{c} = 1.5$ and $m_{b} = 4.75$ GeV for charm and bottom, respectively. The uncertainties of the predictions are estimated by changing quark mass by $\pm 0.25$ GeV and by varying scales by a factor $2$.
The gray shaded bands drawn in the following figures represent these both sources of uncertainties summed in quadrature. The MSTW08LO \cite{Martin:2009iq} collinear parton distribution function (PDF) is used for the calculation of the KMR unintegrated gluon density.

The transition from quarks and gluons to hadrons, called hadronization or parton fragmentation, can be so far approached only through phenomenological models. In principle, in the case of multi-particle final states the Lund string model \cite{Andersson:1983ia} and the cluster fragmentation model \cite{Webber:1983if} are often used. However, the hadronization of heavy quarks in non-Monte-Carlo calculations is usually done with the help of fragmentation functions (FF). The latter are similar objects as the parton distribution functions (PDFs) and provide the probability for finding a hadron produced from a high energy quark or gluon.

Considering fragmentation of a high energy quark (parton) $q$ with zero transverse momentum $p_t$ into a hadron $q \to h + X$ one usually assumes that the transition is soft and does not add any transverse momentum. In consequence it is a delta function in transverse momentum.

Defining $D(z)dz$ as the probability for the quark $q$ fragmenting into a hadron $h$ which carries a fraction $z$ of its longitudinal momentum, the spectrum of hadrons is given by
\begin{equation}
\frac{d\sigma_h}{dx_h d^2 p_{t,h}} = \delta^{(2)}(\vec{p}_{t,h}) \int dz dx D(z) \frac{d\sigma_q}{dx}\delta(x_h - zx) \;
\end{equation}
or 
\begin{equation}
\frac{d\sigma_h}{dx_h} = \int \frac{dz}{z} D(z) \frac{d\sigma_q}{dx}\Bigg\vert_{x = x_h/z} \;.
\end{equation}

This can be generalized to the fragmentation at finite $p_t$ with intrinsic transverse momentum $\kappa_t$. Thus the incoming quark has transverse momentum $p_{t,q}$ and the outgoing hadron transverse momentum can be decomposed as $p_{t,h} = \kappa_t + z p_{t,q}$. Moving back to the situation where $ p_{t,q} = 0$ one can argue that the intrinsic transverse momentum $\kappa_t$ must be small.

Introducing the transverse momentum dependent fragmentation probablity $D(z,\kappa_t)dzd\kappa_t$ one obtains
the hadron spectum as
\begin{equation}
\frac{d\sigma_h}{dx_h d^2 p_{t,h}} = \int dz d^2\kappa_t D(z,\kappa_t) dx_q d^2 p_{t,q} \frac{d\sigma_q}{dx_q d^2 p_{t,q}}
\delta(x_h - zx_q) \delta^{(2)}(\vec{p}_{t,h} - \vec{\kappa_t} - z\vec{p}_{t,q}) \;.
\end{equation}
However, one often neglects the intrinsic transverse momentum assuming:
\begin{equation}
D(z,\kappa_t) = D(z)\delta^{(2)}(\vec{\kappa_t}).
\end{equation}
Then the general formula reads
\begin{equation}
\frac{d\sigma_h}{dx_h d^2 p_{t,h}} = \int dz d x_q d^2 p_{t,q} D(z) \frac{d\sigma_q}{dx_q d^2 p_{t,q}}
\delta(x_h - zx_q) \delta^{(2)}(\vec{p}_{t,h} - z \vec{p}_{t,q}) \;,
\end{equation}
or integrating out the $\delta$-functions
\begin{equation}
\frac{d\sigma_h}{dx_h d^2 p_{t,h}} = \int \frac{dz}{z^2} D(z) \frac{d\sigma_q}{dx_q d^2 p_{t,q}}\Bigg\vert_{x_q = x_h/z \atop p_{t,q} = p_{t,h}/z } \;.
\end{equation}
It is belived that this procedure provides correct implementation of small intrinsic transverse momentum into the splitting. Since the hadron on-shell four momentum is fully specified by $x_h$ and $p_{t,h}$, starting from the above formula one can calculate many different distributions. Especially conversion to the rapidity distributions is very simple because of the trivial jacobian:
\begin{equation}
\frac{xd\sigma}{dx d^2 p_t} = \frac{d\sigma}{dy d^2 p_t}.
\end{equation}
It can be further written
\begin{eqnarray}
\frac{d\sigma_h}{dy_h d^2 p_{t,h}}\!&=&\!\frac{x_h d\sigma_h}{dx_h d^2 p_{t,h}} = \int dz dx_q d^2 p_{t,q} D(z) \frac{d\sigma_q}{dx_q d^2 p_{t,q}}
x_h \cdot \delta(x_h - z x_q) \delta^{(2)}(\vec{p}_{t,h} - z \vec{p}_{t,q})\nonumber \\
&=& \int dz \frac{dx_q}{x_q} d^2 p_{t,q} D(z) \frac{x_q d\sigma_q}{dx_q d^2 p_{t,q}}
x_h \cdot \delta(x_h - z x_q) \delta^{(2)}(\vec{p}_{t,h} - z \vec{p}_{t,q}).
\end{eqnarray}
Neglecting masses, one has
\begin{equation}
x_h = \frac{p_{t,h}}{\sqrt{s}} e^{y_h}, \;\;\;\;\;\; x_q = \frac{p_{t,q}}{\sqrt{s}} e^{y_q} \;,
\end{equation}
so that,
\begin{eqnarray}
\frac{d\sigma_h}{dy_h d^2 p_{t,h}}\!&=&\!\frac{x_h d\sigma_h}{dx_h d^2 p_{t,h}} = \int dz dy_q d^2 p_{t,q} D(z) \frac{d\sigma_q}{dy_q d^2 p_{t,q}}
\delta(1 - e^{(y_q - y_h)}) \delta^{(2)}(\vec{p}_{t,h} - z \vec{p}_{t,q})\nonumber \\
&=& \int dz dy_q d^2 p_{t,q} D(z) \frac{d\sigma_q}{dy_q d^2 p_{t,q}}
\delta(y_q - y_h) \delta^{(2)}(\vec{p}_{t,h} - z \vec{p}_{t,q}) \nonumber \\
&=& \int \frac{dz}{z^2} D(z) \frac{d\sigma_q}{dy_q d^2 p_{t,q}} \Bigg\vert_{y_q = y_h \atop p_{t,q} = p_{t,h}/z} \;.
\end{eqnarray}
This way one gets (reproduces) the standard formula for the fragmentation in the case of light hadrons.

When going to more general case, taking masses into account and introducing $m_t = \sqrt{p_t^2 + m^2}$, one gets
\begin{equation}
x_h = \frac{m_t^h}{\sqrt{s}} e^{y_h}, \;\;\;\;\;\; x_q = \frac{m_t^q}{\sqrt{s}} e^{y_q} \;,
\end{equation}
or
\begin{equation}
y_h = \log \left( \frac{x_h\sqrt{s}}{m_t^h} \right), \;\;\;\;\;\; y_q = \log \left( \frac{x_q\sqrt{s}}{m_t^q} \right), 
\end{equation}
then, the $z$-dependent rapidity shift between quark and hadron reads
\begin{equation}
\delta y = y_q - y_h = \log \left( \frac{x_q m_t^h}{x_h m_t^q} \right) = \log \left( \frac{m_t^h}{z m_t^q} \right).
\end{equation}
The delta function now becomes
\begin{eqnarray}
x_h \delta(x_h - x_q z) &=& \delta \left( 1 - \frac{z x_q}{x_h} \right)= \delta \left( 1- \frac{z m_t^q}{m_t^h} e^{(y_q - y_h)} \right) \nonumber \\
&=&
\delta \left( 1 - e^{(y^q - \delta y - y_h)} \right) = \delta(y_q - \delta y - y^h).
\end{eqnarray}
Finally,
\begin{eqnarray}
\frac{d\sigma_h}{dy_h d^2 p_{t,h}} &=& \int dz dy_q d^2 p_{t,q} D(z) \frac{d\sigma_q}{dy_q d^2 p_{t,q}}
\delta(y_h - y_q + \delta y)\delta^{(2)}(\vec{p}_{t,h} - z \vec{p}_{t,q}) \nonumber \\
&=& \int \frac{dz}{z^2} D(z) \frac{d\sigma_q}{dy_q d^2 p_{t,q}}\Bigg\vert_{y_q = y_h + \delta y \atop p_{t,q} = p_{t,h}/z} \;. 
\end{eqnarray}
In turn, using $p_{t,q} = p_{t,h}/z$, the rapidity shift $\delta y$ can be rewritten
\begin{equation}
\delta y = \frac{1}{2}\log \left( \frac{p^2_{t,h} + m_h^2}{p^2_{t,h} + z^2 m_q^2} \right).
\end{equation}

It is clear that the rescalling of the transverse momentum is the most important effect. This is because one deals with very steep functions of transverse momenta. From the reason that rapidity spectra are usually flat, or slowly varying, the shift $\delta y$ is not so important. In fact, it is entirely negligible, if $p^2_{t,h} \gg m_q^2, m_h^2$. The shift is most important at very small $p^2_{t,h} \ll m_q^2, m_h^2$, where it becomes
\begin{equation}
\delta y \sim \log\left( \frac{m_h}{zm_q}  \right) \approx \log\left( \frac{1}{z} \right). 
\end{equation}
It is worth to notice, that at finite $p_{t,h}$ it should never be really large: small $z$ is damped by the fact that the quark spectrum drops rapidly as a function of $p_{t,q} = p_{t,h}/z$. However, at $p_{t,h}$ = 0, that suppression causes an effect and the whole integral over $z$ becomes important, with very small $z$ causing large rapidity shifts. Fortunately, for heavy quarks, the fragmentation function is peaked at large $z$ (see e.g. \cite{Luszczak:2014cxa}). Moreover, one has to remember, that taking into account the small-$z$ region in the fragmentation function is theoretically not warranted, since the standard DGLAP approach to fragmentation breaks down in this region. 

Taking all together, according to the above formalism, in the following numerical calculations
the inclusive distributions of open charm and bottom hadrons $h =D, B$ are obtained through a convolution of inclusive distributions of heavy quarks/antiquarks and $Q \to h$ fragmentation functions:
\begin{equation}
\frac{d \sigma(pp \rightarrow h \bar{h} X)}{d y_h d^2 p_{t,h}} \approx
\int_0^1 \frac{dz}{z^2} D_{Q \to h}(z)
\frac{d \sigma(pp \rightarrow Q \overline{Q} X)}{d y_Q d^2 p_{t,Q}}
\Bigg\vert_{y_Q = y_h \atop p_{t,Q} = p_{t,h}/z} \;,
\label{Q_to_h}
\end{equation}
where $p_{t,Q} = \frac{p_{t,h}}{z}$ and $z$ is the fraction of
longitudinal momentum of heavy quark $Q$ carried by a hadron $h$.
The origin why the approximation typical for light hadrons assuming that $y_{Q}$ is
unchanged in the fragmentation process, i.e. $y_h = y_Q$, is also applied in the case of heavy hadrons was carefully clarified in the previous paragraph and is commonly accepted. 

As a default set in all the following numerical calculations the standard Peterson model of fragmentation function \cite{Peterson:1982ak} with the parameters $\varepsilon_{c} = 0.02$ and $\varepsilon_{b} = 0.001$ is applied. This choice of fragmentation function and parameters is based on our previous theoretical studies of open charm production at the LHC \cite{Maciula:2013wg}, where detailed analysis of uncertainties related with application of different models of FFs was done. Here, we decided not to repeat all the previously analyzed issues and take into consideration only the most data-favoured scenario\footnote{This is also consistent with prescription applied in the FONNL framework, where rather harder fragmentation functions are suggested \cite{Cacciari:2005rk}.}. The main conclusions should not change when moving from LHC to RHIC energies and the uncertainties due to the fragmentation effects may be neglected with respect to those related to the perturbative part of the calculation. 
  
In the calculations of the cross sections for $D^{0}$ and $D^{*}$ mesons the fragmentation functions should be normalized to the relevant branching fractions $\textrm{BR}(c \to D)$, e.g. from Ref.~\cite{Lohrmann:2011np}. However, the measured by STAR differential distributions for $D^{0}$ and $D^{*}$ meson are normalized to the parton-level $c\bar c$ cross section which simply means that the $\textrm{BR}(c \to D) = 1$ should be taken in numerical calculations.

Theoretical predictions for production of non-photonic leptons in proton-proton scattering is a three-step process. The whole procedure can be written in the following schematic way:
\begin{equation}
\frac{d \sigma(pp \to l^{\pm} X )}{d y_e d^2 p_{t,e}} =
\frac{d \sigma(pp \to Q X)}{d y_Q d^2 p_{t,Q}} \otimes
D_{Q \to h} \otimes
f_{h \to l^{\pm}} \; ,
\label{whole_procedure}
\end{equation}
where the symbol $\otimes$ denotes a generic convolution. Thus, the cross section for non-photonic leptons is a convolution of the cross section for heavy quarks with fragmentation function $D_{Q \to h}$ and with semileptonic decay function
$f_{h \to l^{\pm}}$ for heavy mesons.

In principle, the semileptonic decays can be calculated \cite{Artuso:2008vf}.
The simplest approach to describe the decays of $D$ and $B$ mesons is given by the spectator model \cite{Artuso:2008vf}, where the QCD effects from the higher-order corrections between heavy $Q$ and light $q$ quarks are neglected. This model works better for bottom quarks since there the mass sufficiently suppresses these corrections. In the case of charm, the QCD effects become more important but they can be also qualitatively modelled.

Since there are many decay channels with different
number of particles the above procedure is not easy and rather labor-intensive. It introduces some model 
uncertainties and requires inclusion of all final state channels explicitly.

An alternative way to incorporate semileptonic decays into theoretical model is to take relevant experimental input. For example, the CLEO \cite{Adam:2006nu} and BABAR \cite{Aubert:2004td} collaborations have measured very precisely the momentum spectrum of electrons/positrons coming from the decays of $D$ and $B$ mesons, respectively. This is done by producing resonances: $\Psi(3770)$
which decays into $D$ and $\bar D$ mesons, and $\Upsilon(4S)$ which decays into $B$ and $\bar B$ mesons.

This less ambitious but more pragmatic approach is based on purely empirical fits to (not absolutely normalized) CLEO and BABAR experimental data points. These electron decay functions should account for the proper
branching fractions which are known experimentally (see e.g. \cite{Beringer:1900zz,Adam:2006nu,Aubert:2004td}).
The branching fractions for various species of $D$ mesons are different:
\begin{eqnarray}
&&\mathrm{BR}(D^+\to~e^+ \nu_e X)=16.13\pm 0.20(\mathrm{stat.})\pm 0.33(\mathrm{syst.})\%,\nonumber \\
&&\mathrm{BR}(D^0\to~e^+ \nu_e X)=6.46\pm 0.17(\mathrm{stat.})\pm 0.13(\mathrm{syst.})\%. 
\end{eqnarray}
Because the shapes of positron spectra for both decays are 
identical within error bars we can take the average value of BR($D\!\to\!e \, \nu_e \, X) \approx 10 \%$ and 
simplify the calculation. In turn, the branching fraction of open bottom is found to be:
\begin{equation}
\mathrm{BR}(B\to e \, \nu_e \, X) = 10.36 \pm 0.06(\mathrm{stat.}) \pm 0.23(\mathrm{syst.}) \% .
\end{equation} 

After renormalizing to experimental branching fractions the adjusted decay functions are then use to generate leptons 
in the rest frame of the decaying $D$ and $B$ mesons in a Monte Carlo approach. This way one can avoid all uncertainties associated with explicit calculations of semileptonic decays of mesons.

In both cases the heavy mesons are almost at rest, so in practice one measures the meson rest frame
distributions of electrons/positrons. With this assumption one can find a good fit to the CLEO and BABAR data with:
\begin{eqnarray}
f^{Lab}_{CLEO}(p) &=& 12.55 (p+0.02)^{2.55} (0.98-p)^{2.75}\;, \label{CLEO_fit_function} \\
f^{Lab}_{BABAR}(p) &=& \left( 126.16+14293.09 \exp(-2.24 \log(2.51-0.97 p)^2 \right) \nonumber \\
&&\times \left( -41.79+42.78 \exp(-0.5(|p-1.27|) /1.8 )^{8.78} \right )\;.
 \label{BABAR_fit_function}
\end{eqnarray}
In these purely empirical parametrizations $p$ must be taken in GeV.

In order to take into account the small effect of the non-zero motion of the $D$ mesons in 
the case of the CLEO experiment and of the $B$ mesons in the case of the BABAR experiment, the above parametrizations of the fits in the laboratory frames have to be modified. The improvement can be achieved by including the boost of the new modified rest frame functions to the CLEO and BABAR laboratory frames. The quality of fits from Eqs.~(\ref{CLEO_fit_function}) and (\ref{BABAR_fit_function}) will be reproduced. The $D$ and $B$ rest frame decay functions take the following form:
\begin{eqnarray}
f^{Rest}_{CLEO}(p) &=& 12.7 (p+0.047)^{2.72} (0.9-p)^{2.21}\;, \label{CLEO_fit_boost}  \\
f^{Rest}_{BABAR}(p) &=& \left( 126.16+14511.2 \exp(-1.93 \log(2.7-1.0825 p)^2 \right) \nonumber \\
&&\times \left( -41.79+42.78 \exp(-0.5(|p-1.27|) /1.8 )^{8.78} \right )\;.
 \label{BABAR_fit_boost}
\end{eqnarray}
%

\begin{figure}[tb]
\begin{minipage}{0.47\textwidth}
 \centerline{\includegraphics[width=1.0\textwidth]{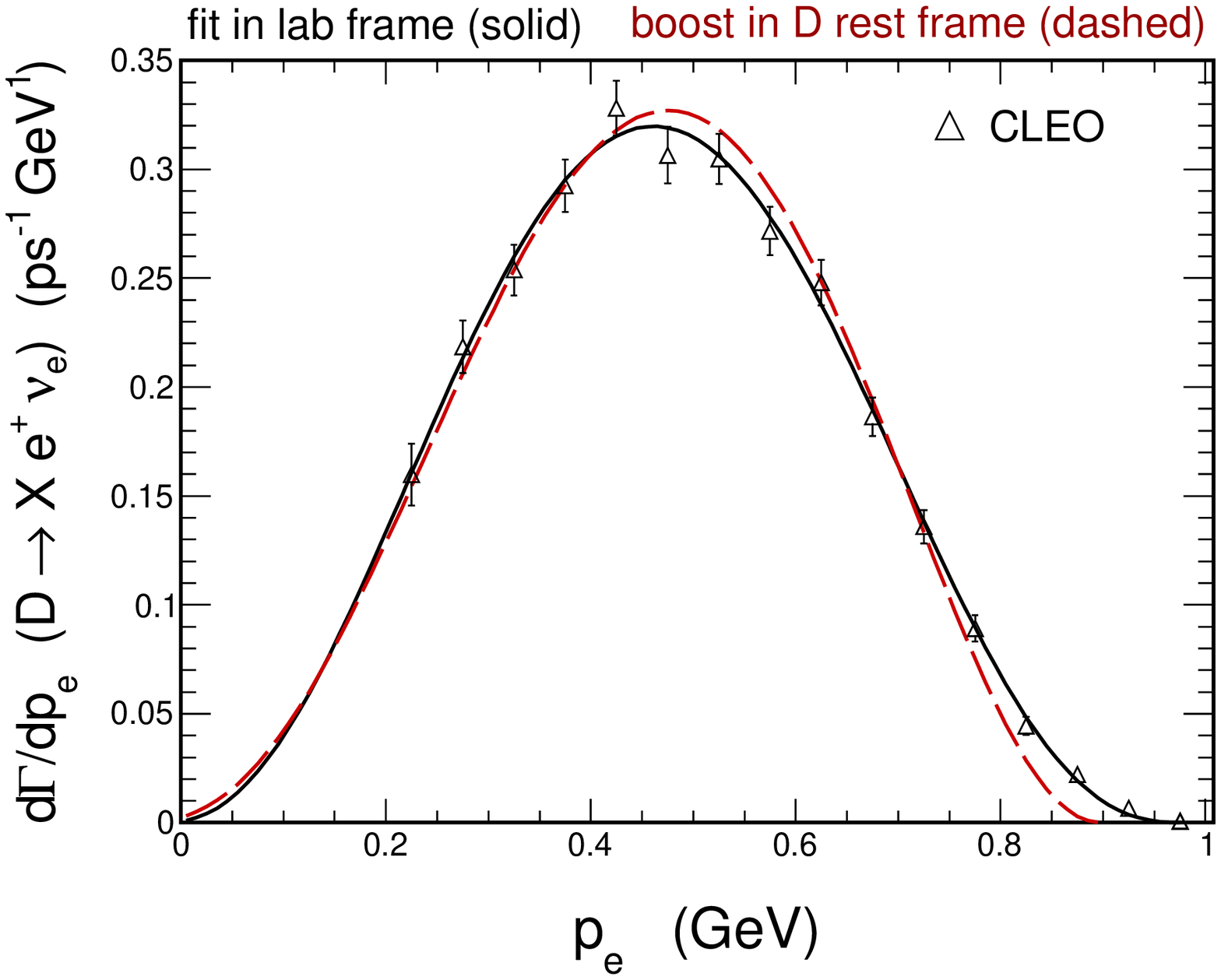}}
\end{minipage}
\hspace{0.5cm}
\begin{minipage}{0.47\textwidth}
 \centerline{\includegraphics[width=1.0\textwidth]{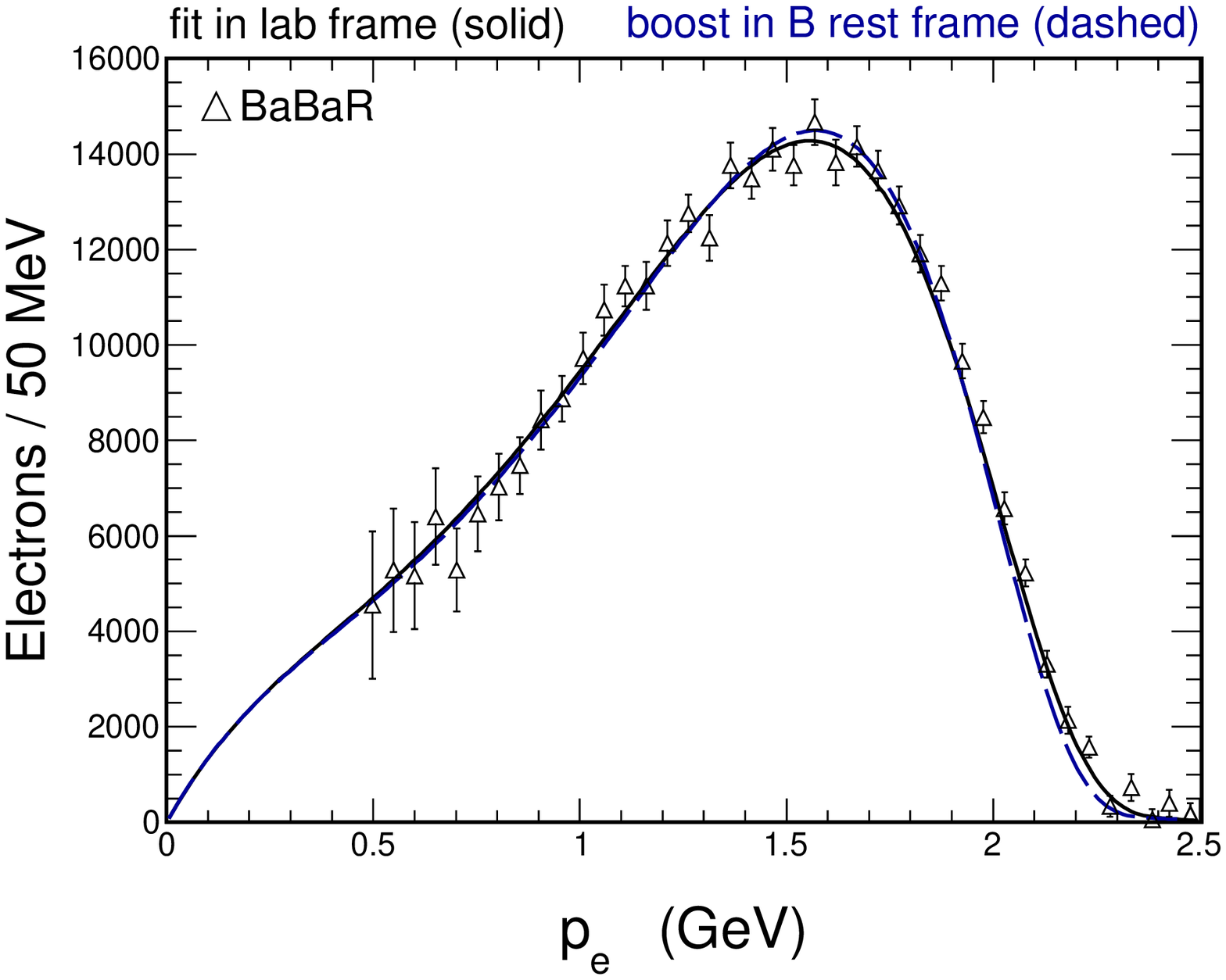}}
\end{minipage}
   \caption{
\small Fits to the CLEO (left) and BABAR (right) data. The solid lines correspond to the parametrizations in the laboratory frames and the dashed lines to the meson rest frames, which represent incorporation of effects related to the non-zero motion of decaying mesons.}
 \label{fig:p-decay-1}
\end{figure}

Both, laboratory and rest frame parametrizations of the semileptonic decay functions for $D$ and $B$ mesons are drawn in Fig.~\ref{fig:p-decay-1} together with the CLEO (left panel) and BABAR (right panel) experimental data. Some small differences between the different parametrizations appear only at larger values of electron momentum. The influence of this effect on differential cross sections of non-photonic leptons is expected to be negligible and will be shown when presenting numerical results. Our analytical formulas for the rest frame decay functions only slightly differ from those obtained in Ref.~\cite{Bolzoni:2012kx}.

The theoretical model for non-photonic lepton production in hadronic reactions described here has been recently found to give a very good description of the experimental data collected with the ALICE detector at the LHC \cite{Abelev:2014hla,Abelev:2014gla}.

\section{Numerical results}

The total cross section for charm production in $pp$ scattering extracted from the STAR measurement of $D^{0}$ and $D^{*}$ mesons at $\sqrt{s} = 200$ GeV is $\sigma_{c\bar{c}} = 797\pm210^{208}_{295}$ $\mu$b. Corresponding calculated total cross sections is $\sigma_{c\bar{c}}^{\mathrm{setB}0} = 541$ with the CCFM Jung set$B0$ UGDF. The calculated value is consistent with the measured value taking into account large experimental uncertainties. The total cross section at $\sqrt{s} = 500$ GeV using the same UGDF is predicted to be $\sigma_{c\bar{c}}^{\mathrm{setB}0} = 1006$ $\mu$b.

The STAR collaboration also carried out a measurement of charm production cross sections at midrapidity $\frac{d\sigma_{c\bar{c}}}{dy}|_{y = 0}$. Comparison of the experimental results and the theoretical ones is presented in Table \ref{tableRHIC-Dmesons}.  
Here again the CCFM Jung set$B0$ UGDF give results consistent with the measurements.

\begin{table}[tb]%
\caption{The midrapidity $\frac{d\sigma_{c\bar{c}}}{dy}|_{y = 0}$ cross section for charm production in proton-proton scattering at $\sqrt{s} = 200$ and $500$ GeV: the STAR results versus results of calculation with the Jung set$B0$ UGDF.}
\newcolumntype{Z}{>{\centering\arraybackslash}X}
\label{tableRHIC-Dmesons}
\centering %
\begin{tabularx}{1.0\textwidth}{ZZZZ}
\toprule[0.1em] %
\\[-3.4ex] 
\multicolumn{2}{c}{Experiment: STAR, $\frac{d\sigma_{c\bar{c}}}{dy}|_{y = 0}$}  & Theory: Jung set$B0$ \\ [+0.4ex]
   
\bottomrule[0.1em]
     & \\ [-3.0ex]
 $\sqrt{s} = 200$ GeV  &  $170\pm45^{+38}_{-59}$ $\mu$b  & $130$ $\mu$b \\ [+0.4ex]
 $\sqrt{s} = 500$ GeV  &  $217\pm86\pm73$ $\mu$b         & $191$ $\mu$b \\ [+1.4ex]

\bottomrule[0.1em]

\end{tabularx}

\end{table}

\subsection{Open charm mesons}

\begin{figure}[!h]
\begin{minipage}{0.47\textwidth}
 \centerline{\includegraphics[width=1.0\textwidth]{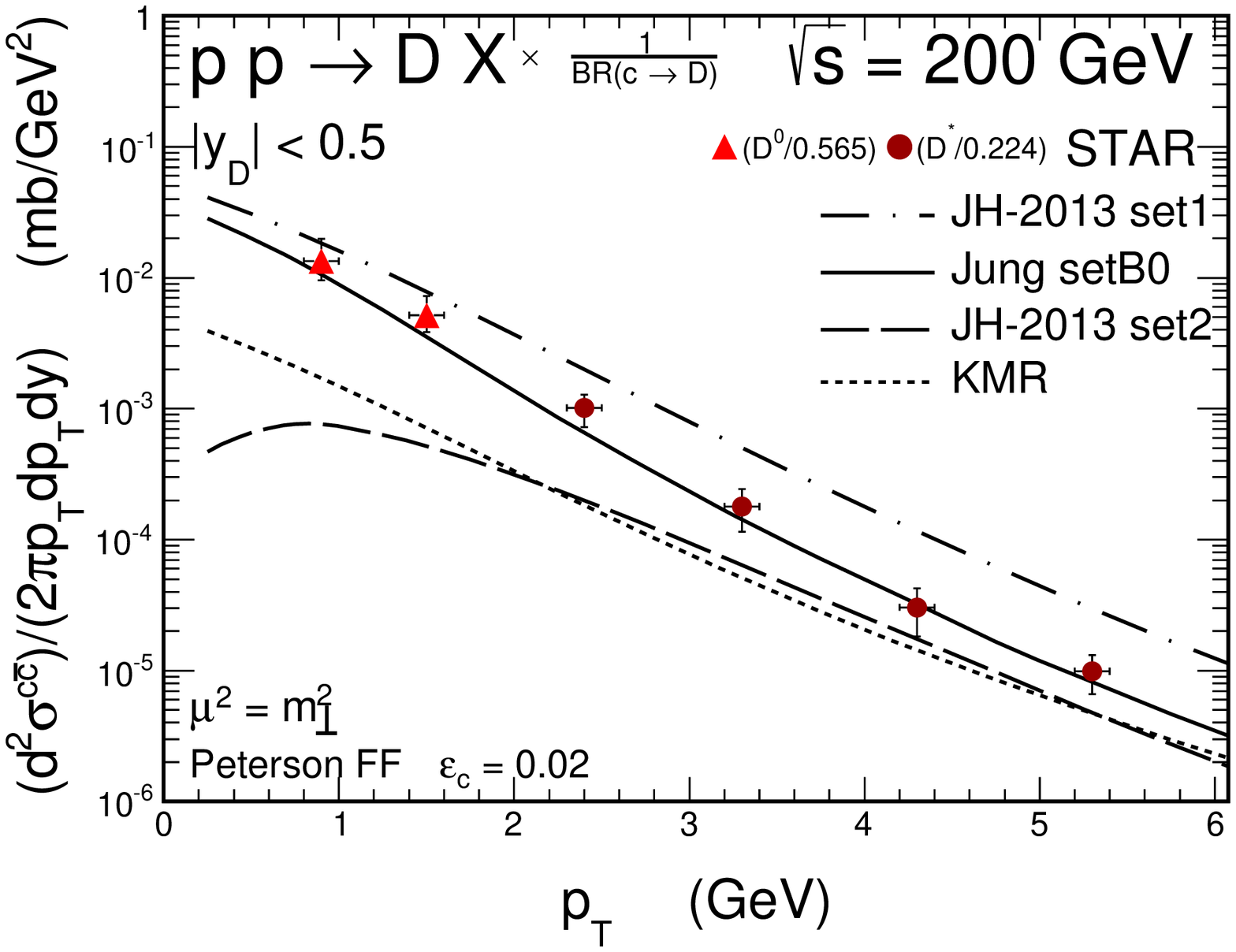}}
\end{minipage}
\hspace{0.5cm}
\begin{minipage}{0.47\textwidth}
 \centerline{\includegraphics[width=1.0\textwidth]{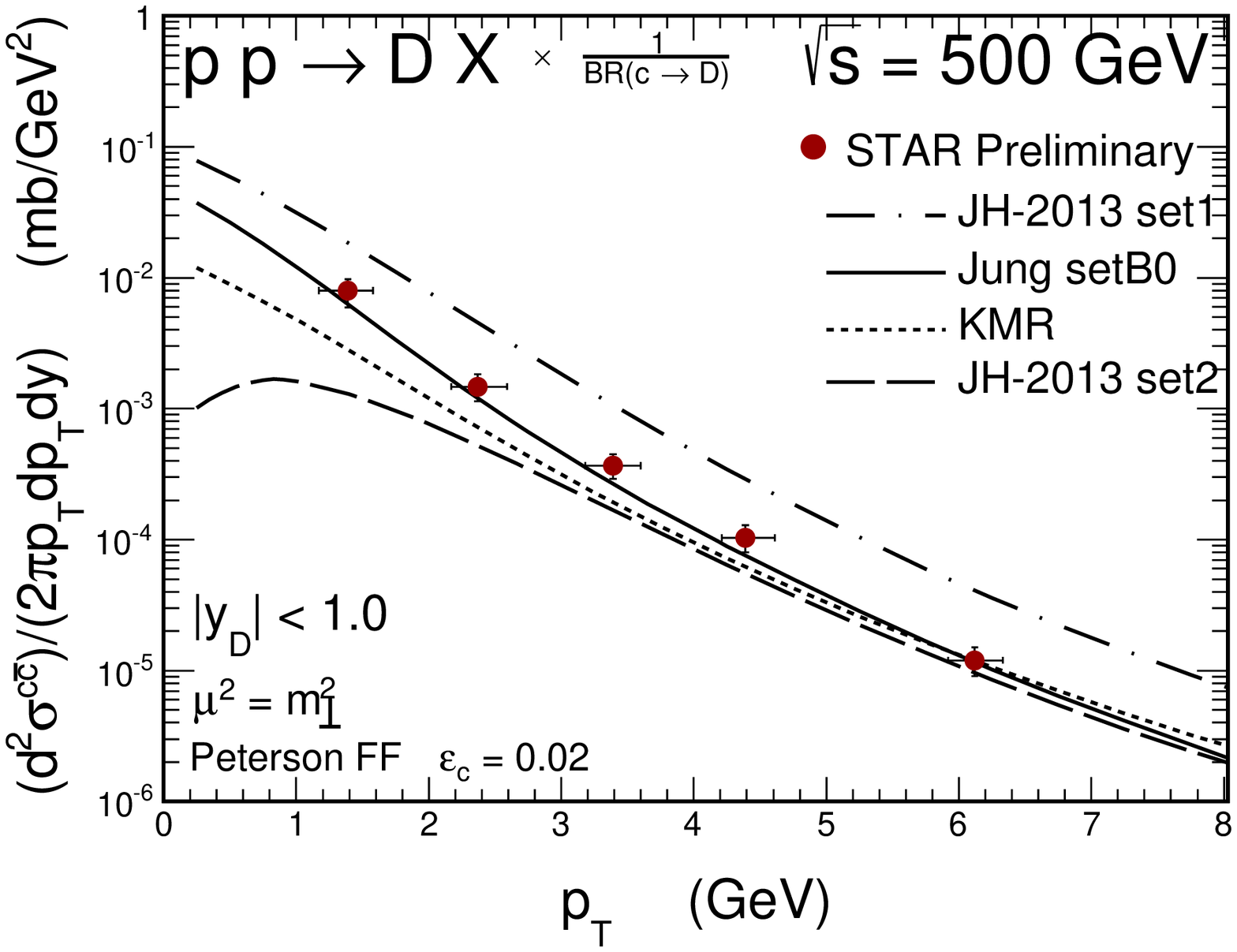}}
\end{minipage}
   \caption{
\small The transverse momentum distribution of $D$ mesons normalized to the parton-level cross section at $\sqrt{s} = 200$ (left) and $500$ GeV (right). The STAR experimental data points are compared to the results of the $k_{t}$-factorization calculations with the KMR (dotted line), the JH2013 set1 (long-dashed-dotted line), set2 (long-dashed line) and the Jung setB$0$ (solid line) UGDFs. }
 \label{fig:pt-star-D1}
\end{figure}

Figure \ref{fig:pt-star-D1} presents transverse momentum distributions of charmed mesons in proton-proton collisions at $\sqrt{s} = 200$ GeV for $|y_{D}| < 0.5$ (left panel) and at $\sqrt{s} = 500$ GeV for $|y_{D}| < 1.0$ (right panel) together with the STAR data points. Both experimental data and theoretical results are normalized to the $c\bar c$ parton-level cross section dividing by the $c \to D^*, D^0$ fragmentation fractions. Results of numerical calculations obtained with the KMR (dotted line), the JH2013 set1 (long-dashed-dotted line), set2 (long-dashed line) and the Jung setB$0$ (solid line) UGDFs are shown. At both energies very good description of the experimental data is obtained with the Jung setB$0$ UGDF. Results calculated with the JH2013 set1 UGDF overestimate the data points in the whole range of measured $p_t$'s.
The JH2013 set2 and the KMR UGDFs significantly underestimate the distribution measured at $\sqrt{s} = 200$ GeV and the situation is only slightly improved in the case of the higher energy where the results of both of them reach the two last data points at larger transverse momenta.    

\begin{figure}[!h]
\begin{minipage}{0.47\textwidth}
 \centerline{\includegraphics[width=1.0\textwidth]{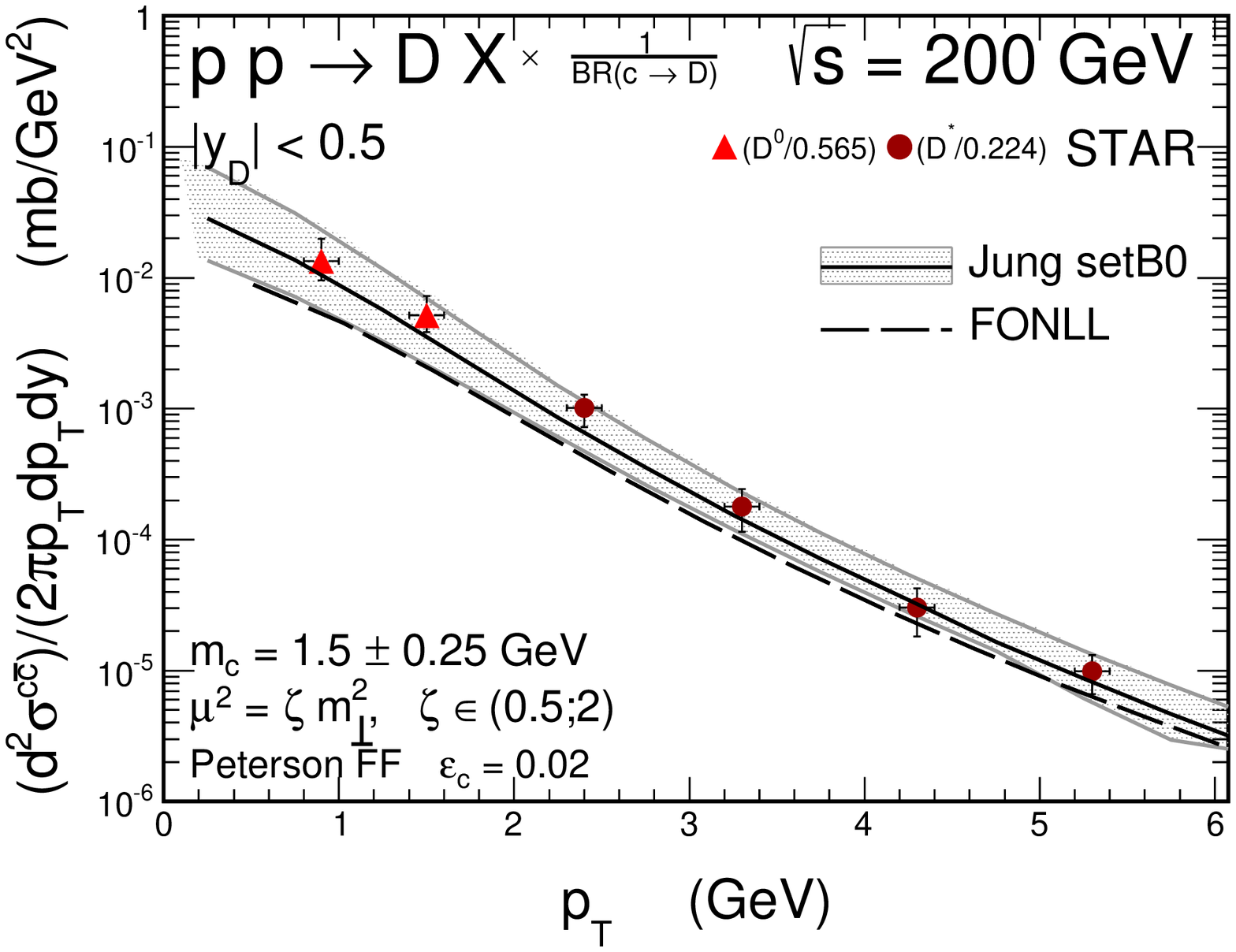}}
\end{minipage}
\hspace{0.5cm}
\begin{minipage}{0.47\textwidth}
 \centerline{\includegraphics[width=1.0\textwidth]{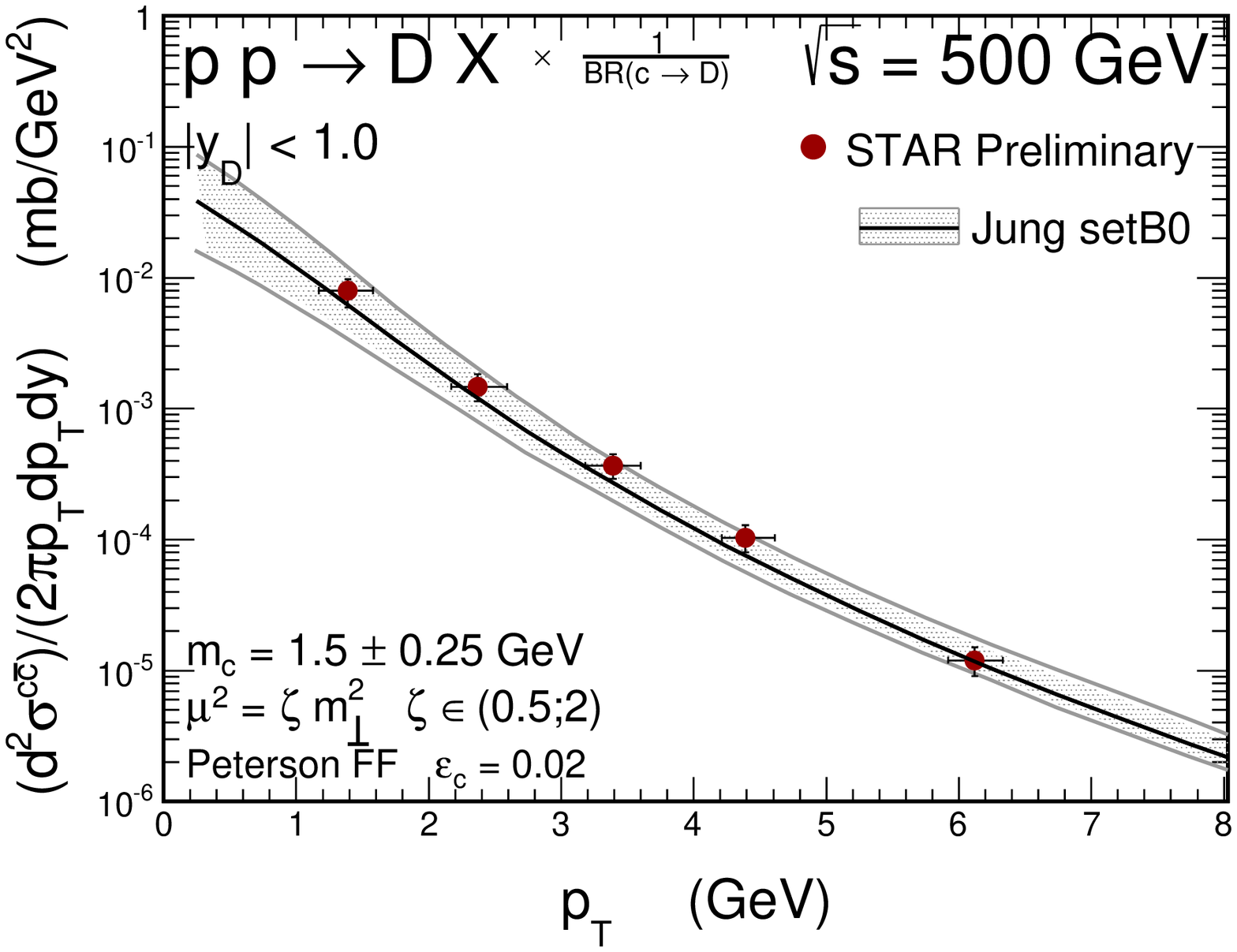}}
\end{minipage}
   \caption{
\small The uncertainties of the theoretical predictions for Jung setB$0$ UGDF at $\sqrt{s} = 200$ (left) and $500$ GeV (right). Uncertainties due to the choice of the factorization and/or renormalization scales and those related to the charm quarks mass are summed in quadrature. In the case of $\sqrt{s} = 200$ GeV data the results obtained within the FONLL framework are drawn for comparison. Details are specified in the plots.}
 \label{fig:pt-D-star-uncert}
\end{figure}

Main uncertainties of the theoretical calculations coming from the perturbative part are shown in Fig.~\ref{fig:pt-D-star-uncert}.
The shaded bands represent uncertainties of the calculations with the Jung setB$0$ UGDF related to the choice of the factorization and/or renormalization scales and those due to the charm quark mass. The result from the FONLL approach is also drawn for comparison. The uncertainties are larger at lower transverse momenta, where the effects of quark mass uncertainties are more important, and decrease with increasing $p_t$. The FONLL predictions underestimate the experimental data almost in the whole measured range. Their central value coincides with the lower-limit of the $k_{t}$-factorization predictions with the Jung setB$0$ UGDF.

\subsection{Non-photonic electrons}

A first theoretical investigation of the non-photonic electron production at RHIC within the framework of the $k_t$-factorization was performed in Ref.~\cite{Luszczak:2008je}. Some missing strength in the description of the measured differential distributions has been reported there, especially in the region of small transverse momenta. In the meantime, the STAR collaboration has published new measurements of non-photonic electrons with separated charm and bottom contributions \cite{Agakishiev:2011mr}. Therefore, it is very interesting to make a revision of theoretical cross sections taking into account the new
results from the hadron-level analysis of charm production at RHIC. Here, our previous results from Ref.~\cite{Luszczak:2008je} are updated by the application of the Jung setB$0$ UGDF which as was shown in the previous subsection works very well for the STAR data on open charm meson production.

The experimental cross sections for charm and bottom flavoured electrons measured at RHIC are collected in Table \ref{table-NPE-RHIC}. The values calculated with the Jung setB$0$ UGDF are consistent with the measurements.

\begin{table}[!h]\small%
\caption{The experimental and theoretical cross sections $\frac{d\sigma}{dy_{e}}|_{y_{e} = 0}$ for non-photonic electron production in proton-proton scattering at $\sqrt{s} = 200$ GeV.}
\newcolumntype{Z}{>{\centering\arraybackslash}X}
\label{table-NPE-RHIC}
\centering %
\begin{tabularx}{1.0\textwidth}{ZZZZ}
\toprule[0.1em] %
\\[-2.4ex] 
                  \multicolumn{2}{c}{Experiment}  & Theory,  Jung setB0 \\ [+0.4ex]
                  
\bottomrule[0.1em]
     & \\ [-2.0ex]
 STAR, $pp\rightarrow c\bar{c} X \rightarrow e X'$                               &  \multirow{2}*{$6.2\pm0.7\pm1.5$ nb}         & \multirow{2}*{$7.55$ nb } \\ [+0.4ex]
        $\frac{d\sigma}{dy_{e}}|_{y_{e} = 0}$,  $3<p_{\perp}<10$ GeV          &                                                 &       \\ [+1.4ex] 
        
  STAR,  $pp\rightarrow b\bar{b} X \rightarrow e X'$                               &  \multirow{2}*{$4.0\pm0.5\pm1.1$ nb}   & \multirow{2}*{$6.65$ nb}    \\ [+0.4ex]       
   $\frac{d\sigma}{dy_{e}}|_{y_{e} = 0}$,  $3<p_{\perp}<10$ GeV                  &                                     &       \\ [+1.4ex]  

 PHENIX, $pp\rightarrow c\bar{c} X \rightarrow e X'$                               &  \multirow{2}*{$5.95\pm0.59\pm2.0$ $\mu$b}         & \multirow{2}*{$5.09$ $\mu$b } \\ [+0.4ex]
        $\frac{d\sigma}{dy_{e}}|_{y_{e} = 0}$,  $p_{\perp}>0.4$ GeV          &                                                 &       \\ [+1.4ex] 

\bottomrule[0.1em]

\end{tabularx}

\end{table}

\begin{figure}[!h]
\begin{minipage}{0.47\textwidth}
 \centerline{\includegraphics[width=1.0\textwidth]{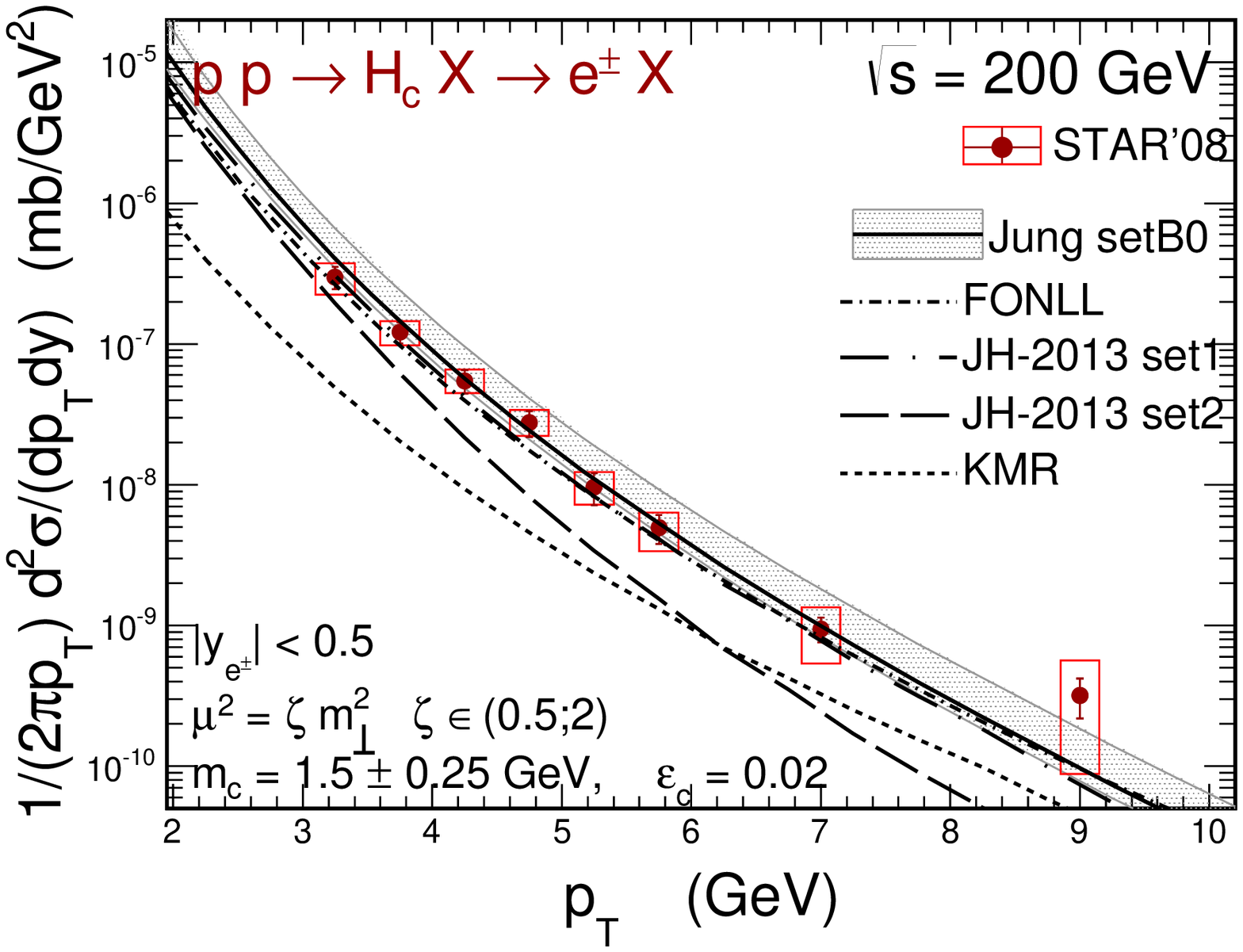}}
\end{minipage}
\hspace{0.5cm}
\begin{minipage}{0.47\textwidth}
 \centerline{\includegraphics[width=1.0\textwidth]{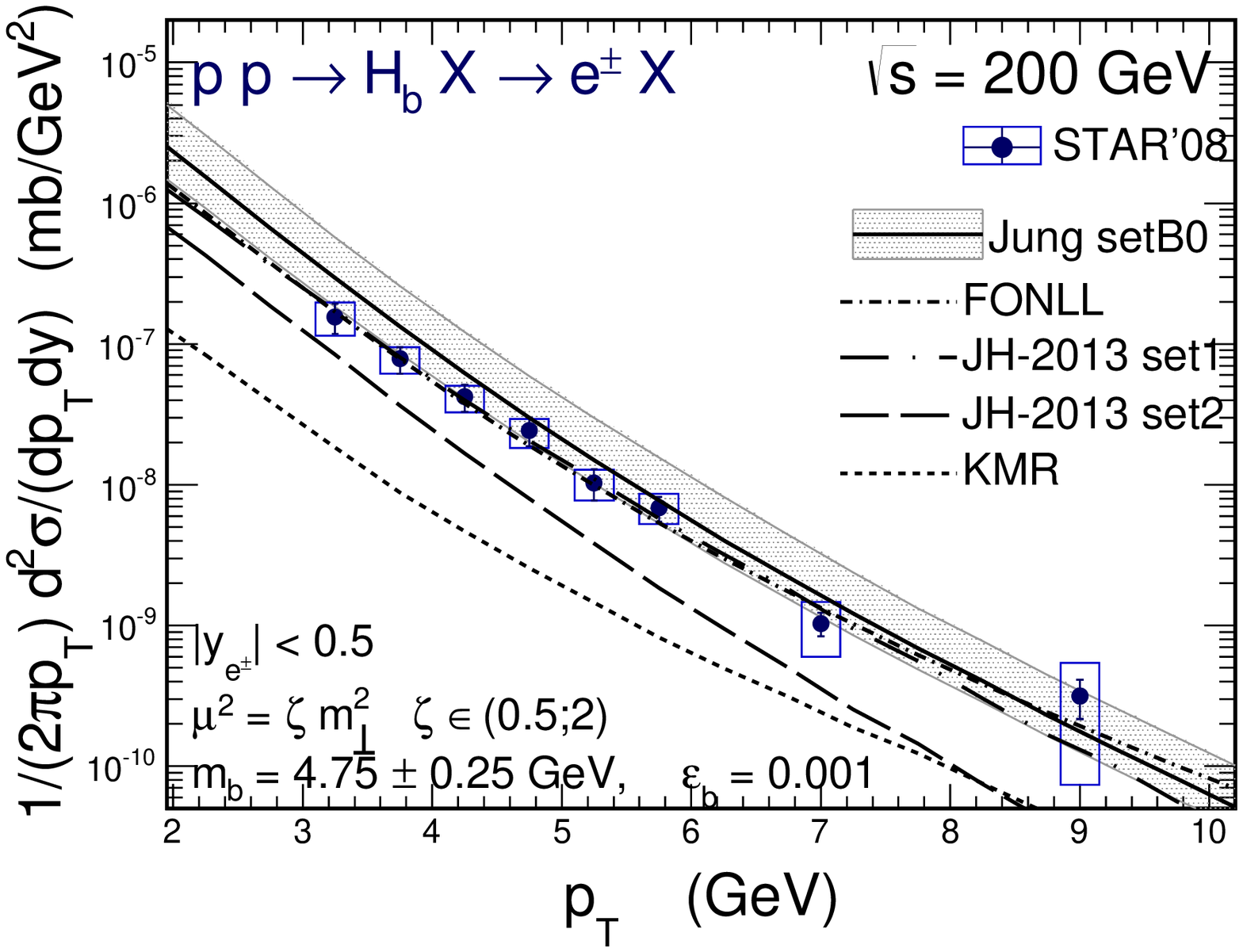}}
\end{minipage}
   \caption{
\small Transverse momentum distributions for electrons from semileptonic decays of charm (left) and bottom hadrons (right) measured in $pp$ scattering at $\sqrt{s} = 200$ GeV. The STAR experimental data are compared to the $k_t$-factorization theoretical predictions obtained with different UGDFs as well as to the FONLL results. Theoretical uncertainties due to quark mass and scales variation are also shown. Further details are specified in the figures. }
 \label{fig:pt-rhic-npe-1}
\end{figure}

Figure~\ref{fig:pt-rhic-npe-1} shows the transverse momentum distributions of electrons from semileptonic decays of charm flavoured hadrons $H_c$ (left panel) and from bottom hadrons $H_b$ (right panel) measured by STAR. The experimental data is compared to the numerical results for the Jung setB$0$ UGDF. The theoretical uncertainties coming from the perturbative part of calculations are also shown for completeness. The rest frame semileptonic decay functions from Eqs.~(\ref{CLEO_fit_boost}) and (\ref{BABAR_fit_boost}) are used. It is also assumed that the charm and bottom baryons decay semileptonically in the same way as $D$ and $B$ mesons, and therefore baryonic contributions may be effectively included by treating the baryons as mesons and taking BR$(c \to D$; $b \to B) = 1$ . The numerical results very well descibe the experimental data. The central value of the Jung setB$0$ UGDF give distributions that are sligthly above the predictions of the FONLL central value, especially in the small-$p_t$ region. In this case also the JH2013 set1 UGDF reasonably describes the data points taking into account experimental uncertainties.
As in the case of open charm data, the lines that corespond to the KMR and JH2013 set2 UGDFs lie much below the measured lepton distributions for both, charm and bottom components.    

\begin{figure}[!h]
\begin{minipage}{0.47\textwidth}
 \centerline{\includegraphics[width=1.0\textwidth]{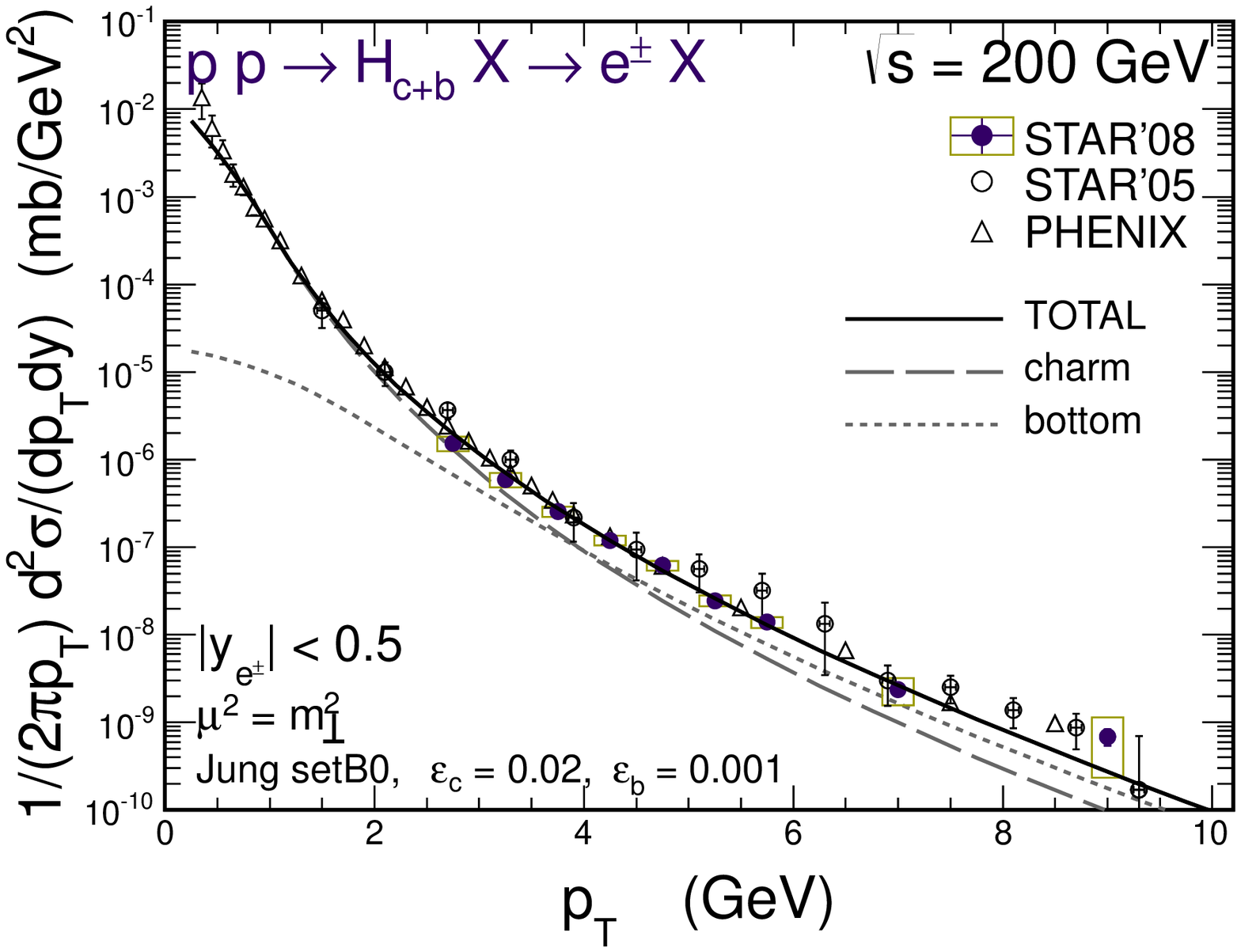}}
\end{minipage}
\hspace{0.5cm}
\begin{minipage}{0.47\textwidth}
 \centerline{\includegraphics[width=1.0\textwidth]{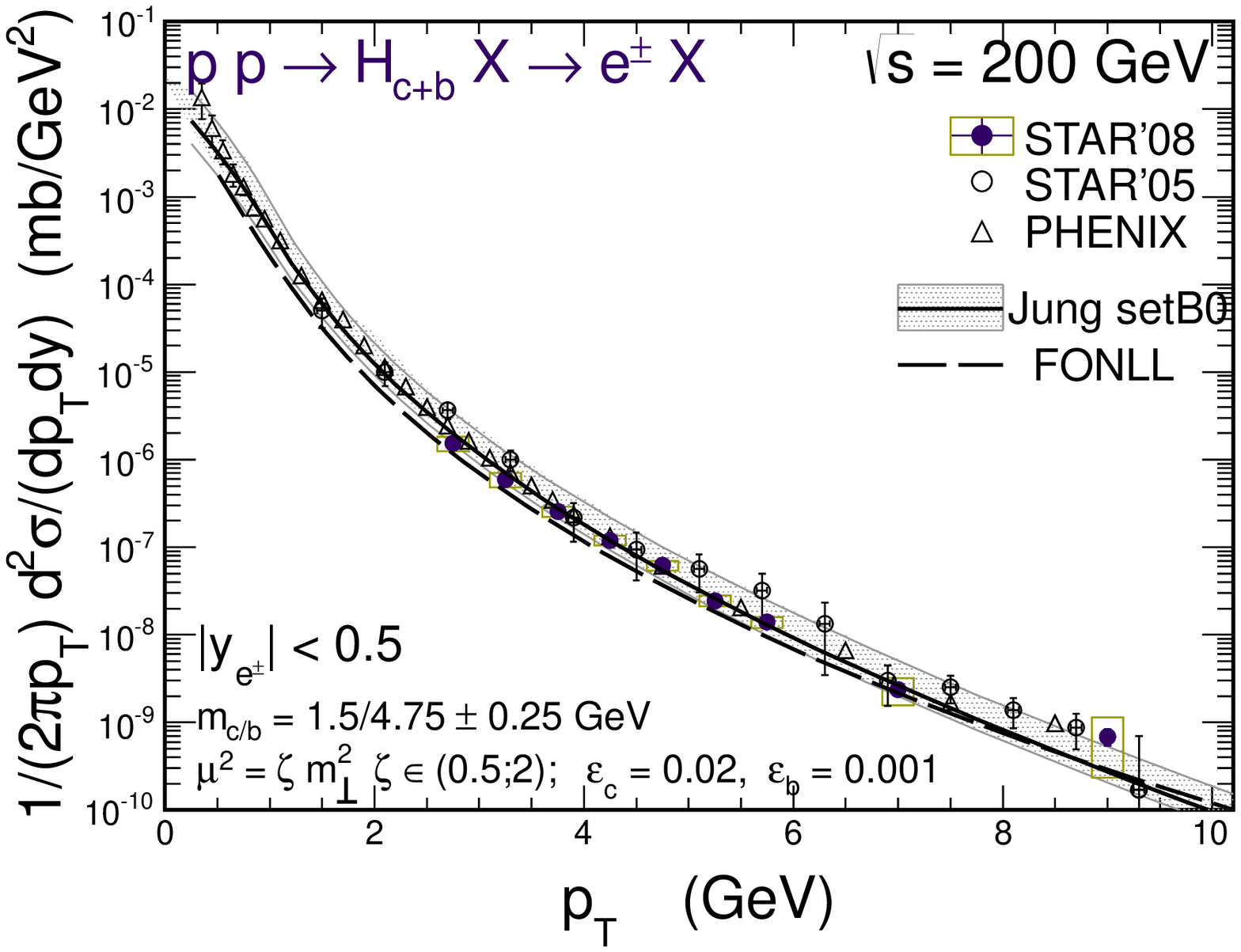}}
\end{minipage}
   \caption{
\small Transverse momentum distributions of electrons coming from both charm and bottom hadrons summed together $H_{c+b}$, measured in $pp$-scattering at $\sqrt{s} = 200$ GeV. The STAR and PHENIX experimental data are compared to the theoretical predictions obtained with the Jung setB$0$ UGDF. Separated charm and bottom contributions (left) and theoretical uncertainties due to quark mass and scales variation (right) are also shown. The FONLL predictions are drawn for comparison. Further details are specified in the figures. }
 \label{fig:pt-rhic-npe-2}
\end{figure}

In Fig.~\ref{fig:pt-rhic-npe-2} the results for summed contributions of charm and bottom flavours are shown.
Here, the results of calculations are compared to the experimental distributions of heavy flavour electrons, that contain both charm and bottom components.
The left panel presents separately charm and bottom contributions as well as the sum of them. In the right panel, the uncertainties of the predictions for the Jung setB$0$ UGDF are drawn together with the lines corresponding to the FONLL results.
In contrast to the previous studies in Ref.~\cite{Luszczak:2008je}, here the $k_t$-factorization results give excellent description of the STAR and PHENIX data, sligthly better than those from the FONLL approach, which are almost identical to the lower for the Jung setB$0$ UGDF. The crossing point between charm and bottom components is found to lies roughly at $p_t = 4$ GeV, which is in agreement with other theoretical investigations (see e.g. Ref.~\cite{Cacciari:2005rk}).

\begin{figure}[!h]
\begin{minipage}{0.47\textwidth}
 \centerline{\includegraphics[width=1.0\textwidth]{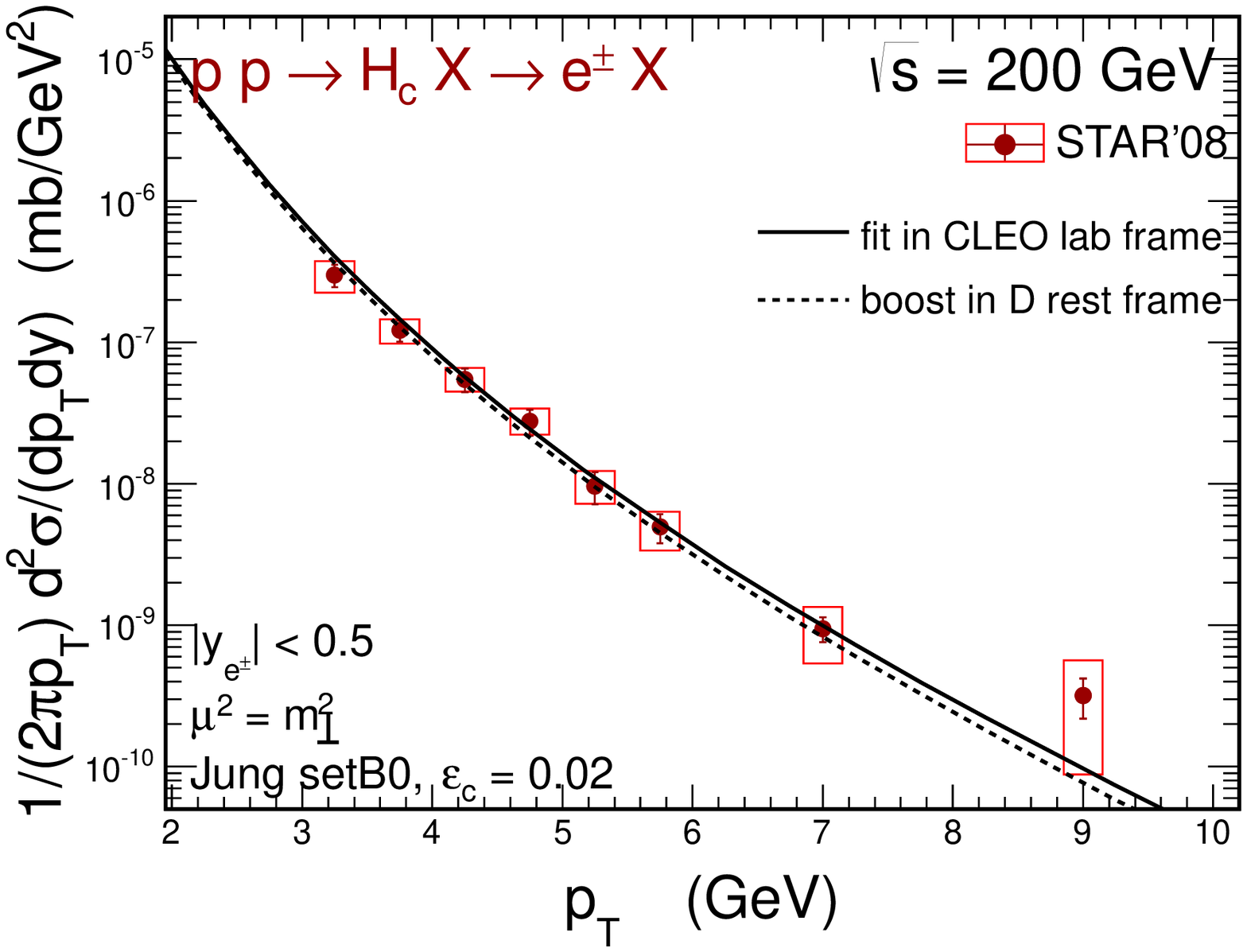}}
\end{minipage}
\hspace{0.5cm}
\begin{minipage}{0.47\textwidth}
 \centerline{\includegraphics[width=1.0\textwidth]{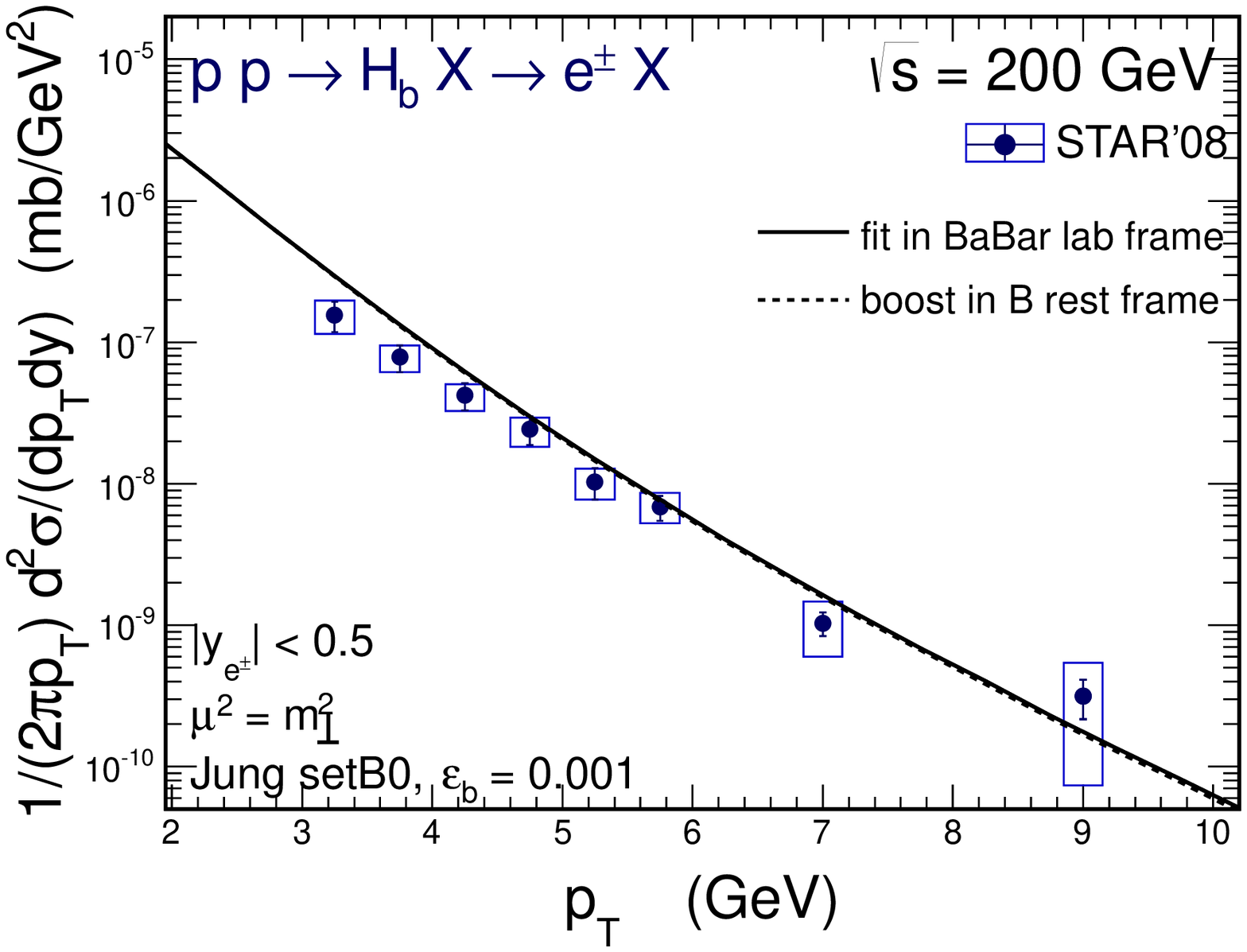}}
\end{minipage}
   \caption{
\small The effect of using two different sets of the semileptonic decays functions, discussed in the present paper, at $\sqrt{s} = 200$ GeV. The results obtained with the laboratory frame fits from Eqs.~(\ref{CLEO_fit_function}) and (\ref{BABAR_fit_function}) are compared to the meson rest frame fits from Eqs.~(\ref{CLEO_fit_boost}) and~(\ref{BABAR_fit_boost}). }
 \label{fig:pt-rhic-npe-ratio-1}
\end{figure}

Finally, the effects related to the fitting procedure of the CLEO and BABAR semileptonic data are depicted in Fig.~\ref{fig:pt-rhic-npe-ratio-1}.
Here, both sets of the semileptonic decay functions are used. As can be observed from the figure these effects do not really affect the electron spectra at RHIC. The difference is very small and sligthtly increases at higher transverse momenta. In the case of charm flavour (left panel), the application of the boosted decay function leads to a damping of the cross section by about $20\%$ at $p_t = 10$ GeV. For the bottom flavour (right panel) the corresponding suppresion is only about $5\%$, which is completely negligible.

\section{Conclusions}

In this paper we have dicussed production of charm mesons and non-photonic
electrons in proton-proton scattering at the BNL RHIC.
The calculation of the charm quark-antiquark pairs has been performed
in the framework of $k_t$-factorization approach which effectively
includes higher-order pQCD corrections.

We have used different models of unintegrated gluon distributions
from the literature, including those that were applied recently to describe
charm data at the LHC and others used to descibe HERA deep-inelastic scattering data.
The hadronization of heavy quarks to mesons has been done by means of fragmentation
function technique. The theoretical transverse momentum distributions of charmed
mesons has been compared with recent experimental data of the STAR collaboration
collected at $\sqrt{s}$ = 200 and 500 GeV.
We have carefully quantified uncertainties related to the choice of factorization/renormalization
scales as well as quark/antiquark masses.
We have obtained very good agreement with the measured cross sections
for the Jung setB$0$ UGDF.
Furthermore, our results have been compared with the results of the FONLL model.
The two approaches give rather similar results.

Semileptonic decays of charmed and bottom mesons have been included
via empirical decay functions fitted to the CLEO and BABAR ($e^+ e^-$) data for vector meson decays.
We have shown that the inclusion of kinematical boost from meson ($D$ or $B$) 
rest frame to the $e^+ e^-$ center of mass (laboratory) system leads to only 
small modifications of the resulting decay functions and as a consequence 
also for the distributions of non-photonic electrons in proton-proton collisions 
at the RHIC energies.
Consequently we have obtained a rather good description of the electron/positron 
transverse momentum distributions of the STAR collaboration with the same UGDF as 
for the charmed mesons.
This also demonstrates indirectly consistency of the meson and 
non-photonic electron data.

\vspace{1cm}

{\bf Acknowledgments}

We are particularly indebted to Jaro Bielcik for a discussion of experimental aspects and to Wolfgang Sch{\"a}fer for a detailed discussion of effects related to quark-to-meson transition. This study was partially supported by the Polish National Science Centre grant DEC-2013/09/D/ST2/03724.


\end{document}